# Magnetization Plateau Observed by Ultra-High Field Faraday Rotation in a Kagomé Antiferromagnet Herbertsmithite

Ryutaro Okuma,[1,2] Daisuke Nakamura,[1] and Shojiro Takeyama[1]
[1]*Institute for Solid State Physics, University of Tokyo, Kashiwanoha, Chiba, 277-8581, Japan*
[2]*Okinawa Institute of Science and Technology Graduate University, Onna-son, Okinawa, 904-0495, Japan*

To capture the high-field magnetization process of herbertsmithite ($ZnCu_3(OH)_6Cl_2$), Faraday rotation (FR) measurements were carried out on a single crystal in magnetic fields of up to 190 T. The magnetization data evaluated from the FR angle exhibited a saturation behavior above 150 T at low temperatures, which was attributed to the 1/3 magnetization plateau. The overall behavior of the magnetization process was reproduced by theoretical models based on the nearest-neighbor Heisenberg model. This suggests that herbertsmithite is a proximate kagomé antiferromagnet hosting an ideal quantum spin liquid in the ground state. A distinguishing feature is the superlinear magnetization increase, which is in contrast to the Brillouin function-type increase observed by conventional magnetization measurements and indicates a reduced contribution from free spins located at the Zn sites to the FR signal.

*Introduction.* Geometrically frustrated magnets are of particular interest owing to the possible realization of various exotic quantum states beyond conventional Néel ordering. The quantum spin liquid has been the holy grail of quantum magnetism since the proposal of a novel nonmagnetic ground state comprising a resonating pattern of singlet pairs in a triangular lattice antiferromagnet by Anderson[1]. A promising platform for realizing a quantum spin liquid is the kagomé lattice, which is a two-dimensional net formed by corner-sharing triangular tiling[2–4]. Theoretical investigations into spin-1/2 kagomé antiferromagnets have arguably revealed the presence of a quantum spin liquid in the ground state[5,6]. Experimentally, spin-1/2 kagomé magnets have been synthesized and some of them show nonmagnetic ground states[7–17].

Among the many candidates for kagomé magnets, herbertsmithite $Zn_xCu_{4-x}(OH)_6Cl_2$ ($x \sim 1$) is recognized as one of the closest candidate for the realization of a spin-1/2 Heisenberg kagomé antiferromagnet[18]. It has an ordered pyrochlore crystal lattice structure as depicted in Fig. 1. The kagomé layer is made up of uniform triangles of $Cu^{2+}$ ions ($3d^9$, $S = 1/2$), while the nonmagnetic triangular layer comprises $Zn^{2+}$ ions partially replaced by $Cu^{2+}$ substitution[19]. The isotropic nearest-neighbor (NN) interaction in the kagomé layer has a strength of 180–300 K, whereas the defect spins in the adjacent triangular layer behave as nearly free spins[20]. Despite such strong interactions, no magnetic order has been observed even at temperatures down to 50 mK[21]. Inelastic neutron scattering and nuclear magnetic resonance have identified gapless or slightly gapped excitations, which are attributed to deconfined spinons excited from the spin liquid ground state[22,23].

However, one important issue remains unresolved, namely, the effects of disorder on the ground state of herbertsmithite. The effects of anti-site disorder in herbertsmithite can be divided into two types: one is dilution of the kagomé plane by $Zn^{2+}$, and the other is occupation of the $Zn^{2+}$ site by Jahn-Teller active $Cu^{2+}$. The former locally breaks the triangles and releases the geometrical frustration. The latter results in the modulation of spin interactions on the kagomé plane. Theoretical studies have suggested that in both cases, a certain amount of randomness induces a nonmagnetic state called a valence bond glass[24] or a random singlet state[25,26] in which singlet pairs form a static spatial pattern rather than the resonating valence bond state. In fact, in the absence of a "smoking gun" for the existence of a quantum spin liquid, all the experimental data reported so far can be reproduced by assuming site-randomness[24–26].

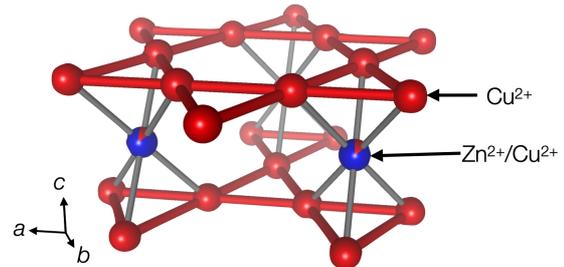

FIG. 1. Magnetic sublattice of herbertsmithite. Red spheres represent $Cu^{2+}$ ions, and red-blue spheres disordered sites with mixed occupancy of $Cu^{2+}$ and $Zn^{2+}$. Red and grey sticks represent the kagomé net intra- and inter-plane bonds with the disordered sites ($Zn^{2+}/Cu^{2+}$), respectively.

Here, we present a unique approach to understanding the effects of disorder on the ground state of herbertsmithite by utilizing the characteristic properties of the kagomé antiferromagnet subjected to very high magnetic fields. In any magnetic materials, a sufficiently high magnetic field forces all the spins to align along the direction of the field. A Bloch wave of spin-flips, called a magnon, is an elementary excitation from such a polarized state. A striking feature of the kagomé antiferromagnet is that the magnon does not propagate but resonates inside a hexagon[27]. Below the saturation field, such hexagonal magnons condense into a series of crystalline states in the absence of quenched disorder[27–33]. Experimentally, these can be observed as 1/3, 5/9, and 7/9 magnetization plateaus on the magnetization curve. As the random spin interactions break the local resonances inside the hexagons in the presence of strong randomness, no crystalline phases appear until the magnetization is saturated. Thus, the existence of magnetization plateaus in a kagomé antiferromagnet is a useful measure of the proximity to the pristine Heisenberg model without random or additional interactions.

It is expected that magnetic fields higher than 100 T are necessary to observe the first (1/3) magnetization plateau in herbertsmithite. Precise measurements using the electro-magnetic induction method, which has been widely used for magnetization measurements in pulsed magnetic fields, are restricted to measurements below 100 T[34,35]. In contrast, the Faraday rotation (FR) technique is promising for more reliable and noiseless measurements in magnetic fields much higher than 100 T[36,37]. The change in the polarization angle $\theta_F$ can be set to be proportional to the magnetization $M$ so that it follows the relation

$$\theta_F = \alpha M d,$$

where α denotes the Verdet constant and $d$ the sample thickness. FR magnetization measurements have been attempted for observing the entire magnetization process in spin 1/2 kagomé antiferromagnets at magnetic fields of up to 160 T[38,39]. These precise FR measurements with reduced noise allowed us to identify the presence of a series of magnetization plateaus in Cd-Kapellasite[39].

The single crystals of herbertsmithite used in this study were synthesized by the hydrothermal transport method[40–42]. The composition ratio $x$ in the $Zn_xCu_{4-x}(OH)_6Cl_2$ crystal was determined to be close to 1 by energy dispersive X-ray spectroscopy. The impurity level of the crystal confirmed by a Magnetic Property Measurement System MPMS-3 (Quantum Design) was similar to that reported for $x = 1.0$[43] as shown in Fig. S1 (the Supplemental Material[44]). The sample was cut along the [101] plane to obtain a hexagonal plate with the dimensions of 1 mm × 1 mm × 0.15 mm. The optically flat crystal was attached to a sapphire substrate and placed into a hand-made miniature liquid-flow type optical cryostat composed entirely of glass-epoxy thin-wall tubes[45]. A magnetic field with a short-pulsed form (pulse width of 7 μs) was generated by a single-turn coil megagauss generator. A magnetic field of up to around 190 T was applied along the direction perpendicular to the [101] plane of the sample. A 532-nm wavelength laser light source was employed in the FR setup. The photon energy corresponds to 2.3 eV, which is at the higher energy tail of the $Cu^{2+}$ $d$-$d$ transition spectrum[46]. The angle $\theta_F$ was calculated from the intensities of the vertical and horizontal components of the transmitted light through the sample, $I_p$ and $I_s$, using the formula $\theta_F = \mathrm{acos}\{(I_p - I_s)/(I_p + I_s)\}$.

*Results and Discussion.* Figure 2 shows the field dependence of the FR signals measured at 6 and 10 K. Temperature pointed out here is the values indicated by the thermocouple placed adjacent to the sample. The sample temperatures could be larger than these values owing to structural restrictions of the optical cryogenic system[36] and the actual temperature should be considered as 6–8 K, and 10–18 K, respectively. The drastic noise increases around 120 T at 6 K and 100 T at 10 K are due to a singularity point of $\theta_F$ at which $I_p$ vanished and resulted in a substantial decrease of the signal-to-noise ratio. Polynomial fitting was performed and the fits were indicated by solid lines to guide the eye in following the noisy data.

At 6 and 10 K, the rotation angle increment was relatively suppressed below 40 T, above which the rotation angle showed a superlinear increase up to 150 T. Interestingly, at the low field region, the Brillouin function-type magnetization increment arising from free spins was absent, a fact that stands in contrast with the results obtained by SQUID measurements. This will be discussed later. The magnetization increment was interrupted by a kink at around $B_p \sim 150$ T and saturated above $B_p$. This indicates the presence of a magnetization plateau with fractional magnetization or the saturation of all spins above $B_p$.

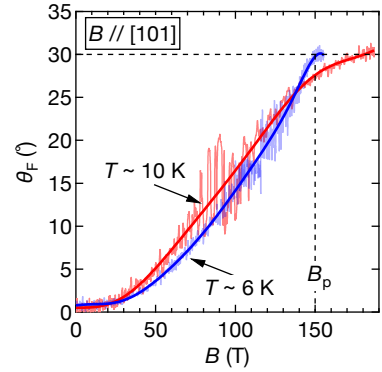

FIG. 2. Evolution of Faraday rotation angle with magnetic field at the temperatures of 6 K (pale blue) and 10 K (pale red). The red and blue solid lines are results of polynomial fitting to the data at 6 and 10 K, respectively. The magnetic field was applied along the [101] axis of the herbertsmithite crystal. The dashed line indicates the position of the magnetization plateau starting from $B_p \sim 150$ T.

We first comment on the discrepancy between our results and the magnetization data obtained by the induction method in magnetic fields up to 60 T generated by a non-destructive long pulse magnet[43]. The latter could be interpreted as the magnetization from the kagomé plane superimposed with a Brillouin function term due to nearly-free interlayer spins. In fact, our sample should show a similar magnetization behavior due to a similar level of impurity concentration.

The discrepancy can be attributed to the site-selectivity of the FR. The FR at the optical wavelength of 532 nm (2.3 eV) arises primarily from the optical transition between the occupied and unoccupied 3$d$ $Cu^{2+}$ orbitals. The optical transition intensity depends on the ligand field splitting and should be sensitive to different chemical environments. Figure 3 illustrates the coordination environment of $Cu^{2+}$ in two different crystallographic sites in herbertsmithite. The copper ion on a kagomé plane has octahedral coordination with strong monoclinic distortion and is surrounded by oxygen and chlorine atoms (Fig. 3a), whereas the copper ion in a defect position has trigonally-distorted octahedral coordination with six surrounding oxygen atoms (Fig.

3b). FR with the photon energy tuned to the ligand field energy splitting of the $Cu^{2+}$ in the kagomé plane is reflected predominantly by the magnetization of the kagomé plane itself and is less contributed from that of the defect site. This fact is confirmed by subtraction of the free spin contribution from the magnetization data[43], which has resulted in a fairly good agreement with that of the Faraday rotation as is demonstrated in Fig. S2 (the Supplemental Material[45]).

The kink around $B_p$ in Fig. 2 evokes the question of whether the magnetization plateau is associated with fractional magnetization or is simply due to the full saturation of the magnetization. The interaction $J$ of herbertsmithite is reported to be around 180–300 K[8,19,20], from which the saturation field is evaluated as $3J = 400$ T–600 T. Therefore, the magnetization kink above $B_p$ cannot be caused by full moment saturation but is most likely due to the 1/3 magnetization plateau. This can also be inferred from the theoretical prediction that the plateau starts to appear above $B \sim J$[28–33,47].

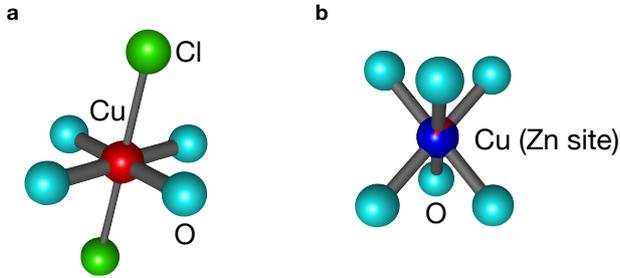

FIG. 3. Local coordination environment of Cu sites. (a) Cu on a kagomé plane, and (b) of a disordered site at an interlayer between kagomé planes.

A comparison of various theoretical models and our experiment results is presented in Fig. 4, in which it was assumed that the FR above $B_p$ is the signal corresponding to 1/3 of the full kagomé plane magnetization at $J = 250$ K. The theoretical calculations correspond to the spin-1/2 NN Heisenberg kagomé antiferromagnet with[26] and without[42,47] disorder. The present FR data match better with the calculated lines modeled without disorder (denoted as PEPS[42] and ED[47] in Fig. 4) than the line with heavy disorder (denoted as EDR[26]). The calculated curve for the latter (EDR) exhibits only a slight anomaly near 1/3 magnetization in contrast to the flat feature observed in the FR signal. The effects of finite temperature and/or existing weak disorder result in the blunt structure at the beginning of the plateau as is observed in the present FR data. The Dzyaloshinskii-Moriya interaction, which is known to exist in this system[48], may also be a possible reason for the blunt structure. of the plateau.

The fact that the present FR data is well reproduced by the magnetization curves calculated by models without disorder has significant implications. The FR data selectively reflects the magnetization from the $Cu^{2+}$ on the kagomé planes and disregards the free spins at the defect sites. By choosing the wavelength of the FR light to match a specific intra-atomic transition, it may be possible to obtain the magnetic information of a magnetic ion in a specific environment.

The discussion above thus leads to the important implication that intrinsic herbertsmithite without free-spin defects, if realized, is well described by the pristine NN Heisenberg kagomé antiferromagnet.

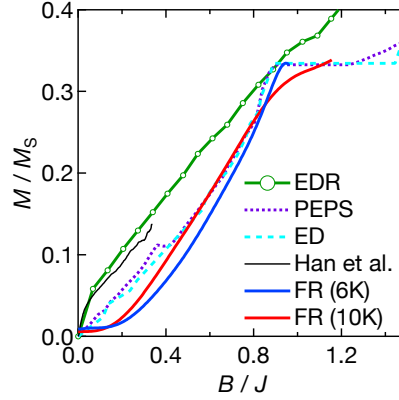

FIG. 4 Comparison of normalized FR data with theoretical calculations: normalized FR data at 6 K (blue) and 10 K (red), magnetization data measured by induction method (thin black)[43], calculation of $S = 1/2$ Heisenberg kagomé antiferromagnet by tensor network (PEPS)[42] (purple dotted line) and exact diagonalization (ED)[47] (dashed sky-blue), and $S = 1/2$ Heisenberg kagomé antiferromagnet with random interaction by exact diagonalization (EDR)[26] (green circle with line). The FR data are normalized by the NN interaction and the FR at the full magnetization, which are assumed to be 250 K and 90°, respectively.

*Conclusion and outlook.* The FR data with a specific optical transition indicates a magnetization curve with the tendency to saturate above $B_p \sim 150$ T below 10 K in the spin-1/2 kagomé antiferromagnet herbertsmithite. The bending feature above $B_p$ was interpreted as the 1/3 magnetization plateau. The overall magnetization features at up to 190 T were well reproduced by the pristine NN Heisenberg kagomé antiferromagnet. These facts indicate that the mineral herbertsmithite, if eliminated from chemical disorder, is close to the ideal Heisenberg kagomé antiferromagnet. However, more compelling evidence for crystallization and magnon sublimation in the kagomé antiferromagnet, e.g., the 5/9 and 7/9 plateaus and an abrupt jump to moment saturation following the 7/9 plateau, remains to be found. The present experiment is anticipated to be extended to around 500 T by using a state-of-the-art electromagnetic flux-compression megagauss generator[49,50] so that the full saturation moment can be observed.

We thank Zenji Hiroi for providing us facilities for crystal growth and fruitful discussion.

# Supplemental Material for "Magnetization Plateau Observed by Ultra-High Field Faraday Rotation in a Kagomé Antiferromagnet Herbertsmithite"

Ryutaro Okuma,[1,2] Daisuke Nakamura,[1] and Shojiro Takeyama[1]

[1]Institute for Solid State Physics, University of Tokyo, Kashiwanoha, Chiba, 277-8581, Japan

[2]Okinawa Institute of Science and Technology Graduate University, Onna-son, Okinawa, 904-0495, Japan

## 1. MAGNETIZATION OF HERBERTSMITHITE AT LOW MAGNETIC FIELDS

The magnetization of herbertsmithite used in this study is measured in magnetic fields up to 7 T and compared with that of the other reports. The similarity in a Brillouin type increase of magnetization implies that amount of free spin of our sample is the same order as that reported in ref. 43.

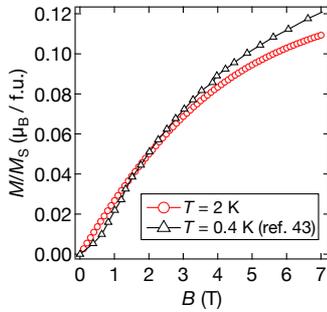

Fig. S1. Magnetization of herbertsmithite in magnetic fields at low magnetic fields. The red circle and black triangle indicate the magnetization process of the sample used in this study at 2 K, and in ref. 43 measured at 0.4 K, respectively.

## 2. BULK MAGNETIZATION AND FARADAY ROTATION ANGLE

Magnetization deduced from Faraday rotation (FR) showed an apparently different feature from that of bulk-magnetization in magnetic fields particularly less than 20 T. This discrepancy is attributed to tuning capability of an optical transition energy involved in FR. FR with a choice of wavelength tuned to the $d$-$d$ intra-atomic optical transition preferentially reflect the magnetization stemming from a magnetic ion in a given ligand field. The incident light wavelength in FR in the main text was set at 532 nm, that is equal to 2.33 eV (18,800 cm$^{-1}$), which corresponds to high energy tail of the major absorption peak (1.44 eV) of the intra-atomic $d$-$d$ transition (the transition of the ligand splitting energy levels; $xz/yz \rightarrow x^2$-$y^2$) [ref.46]. The FR with a laser incident light of 532 nm thus only reflect the magnetization arising from copper ion configured in Fig.3 a. and less from free spins in Fig.3b. This is a major cause of discrepancy between magnetization from FR and the bulk magnetization measured by the conventional induction methods. The Brillouin type increase of magnetization observed in Fig. S1 shows an apparent absence in magnetization deduced from the Faraday rotation. Magnetization of the paramagnetic spin-1/2 is expressed by the Brillouin function,

$$M_{\text{free}}(B) = xg\mu_\text{B} S \tanh\left(\frac{g\mu_\text{B} SB}{k_\text{B}(T + T_{CW})}\right).$$

Here $x$ is an amount of a free spin present in Zn$_{1-x}$Cu$_{3+x}$(OH)$_6$Cl$_2$. $T_{CW}$ denotes the effective temperature arising from the effective antiferromagnetic interaction. The subtraction of free spin contribution $M_{\text{free}}(B)$ from the bulk magnetization (dashed line) is demonstrated in Fig.S2, with appropriate values of $x = 0.15$ and $T_{CW} = 7$ K, which results in a fairly good agreement with data deduced from FR.

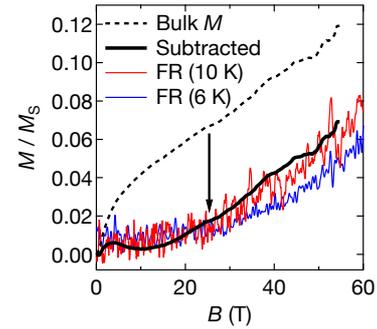

Fig. S2. Comparison of bulk magnetization and that obtained from FR. The red, blue, and black dashed lines indicate the FR data at 6K and 10K, and the subtracted and original bulk magnetization data quoted from ref. 43, respectively.